\documentclass[sigconf,nonacm,screen]{acmart}
\usepackage{url}
\usepackage{xcolor}
\usepackage{graphics}
\usepackage{multirow}
\usepackage{amsmath}

\usepackage{soul}
\newcommand{\ctext}[3][RGB]{%
  \begingroup
  \definecolor{hlcolor}{#1}{#2}\sethlcolor{hlcolor}%
  \hl{#3}%
  \endgroup
}
\newcommand{\CodeAdded}[1]{\ctext[RGB]{226,255,233}{\texttt{#1}}}
\newcommand{\CodeRemoved}[1]{\ctext[RGB]{255,232,230}{\texttt{#1}}}

\usepackage{fixltx2e}

\newcommand{\Heading}[1]{\textbf{#1.}}
\newcommand{\RQ}[1]{\textit{RQ}${}_{\mathrm{#1}}$}
\usepackage{framed}
\newcommand{\Conclusion}[1]{\begin{framed}\noindent #1\end{framed}}

\newcommand{\Refactoring}[1]{\textit{#1}}
\newcommand{\Commit}[1]{\textit{#1}}

\def\detect{\mathit{ref}}
\def\squash{\mathit{sq}}
\def\type{\mathit{type}}
\def\Types{\mathit{types}}
\def\CGR{\mathit{CGR}}
\def\isEffective{\mathit{isEffective}}
\def\Unit{\mathit{unit}}
\def\Frequency{\mathit{Frequency}}
\def\Ratio{\mathit{Ratio}}

\title{Impact of Change Granularity in Refactoring Detection}

\author{Lei Chen and Shinpei Hayashi}
\email{{chenlei,hayashi}@se.c.titech.ac.jp}
\affiliation{%
  \institution{Tokyo Institute of Technology}
  \streetaddress{Ookayama 2-12-1}
  \city{Meguro-ku}
  \state{Tokyo}
  \country{Japan}
  \postcode{152-8550}
}

\begin{abstract}
Detecting refactorings in commit history is essential to improve the comprehension of code changes in code reviews and to provide valuable information for empirical studies on software evolution.
Several techniques have been proposed to detect refactorings accurately at the granularity level of a single commit.
However, refactorings may be performed over multiple commits because of code complexity or other real development problems, which is why attempting to detect refactorings at single-commit granularity is insufficient. 
We observe that some refactorings can be detected only at coarser granularity, that is, changes spread across multiple commits.
Herein, this type of refactoring is referred to as \textit{coarse-grained refactoring} (CGR).
We compared the refactorings detected on different granularities of commits from 19 open-source repositories.
The results show that CGRs are common, and their frequency increases as the granularity becomes coarser.
In addition, we found that \textit{Move}-related refactorings tended to be the most frequent CGRs.
We also analyzed the causes of CGR and suggested that CGRs will be valuable in refactoring research.
\end{abstract}



\begin{document}

\maketitle

\section{Introduction}\label{s:introduction}
Mining refactorings in commit history is essential to help programmers comprehend code changes and code reviews~\cite{kim2012field},and this
can provide valuable information for empirical studies on software evolution\cite{6392107,kim2014empirical}.
For example, Ch\'{a}vez \textit{et al.}~\cite{10.1145/3131151.3131171} and Fernandes \textit{et al.}~\cite{FERNANDES2020106347} detected and analyzed refactorings to investigate the refactoring performance in improving internal quality attributes.

Refactoring detectors~\cite{dig2006automated,10.1145/1882291.1882353,weissgerber-ase2006,prete-icsm2010,silva2017refdiff,Tsantalis:ICSE:2018:RefactoringMiner} detect refactorings by comparing two source code snapshots.
Although traditional approaches aim to detect refactorings over releases~\cite{weissgerber-ase2006,prete-icsm2010}, recent detectors such as RefDiff~\cite{silva2017refdiff,refdiff20} and RefactoringMiner~\cite{Tsantalis:ICSE:2018:RefactoringMiner,Tsantalis:TSE:2020:RefactoringMiner2.0} use a commit as a change unit to detect refactorings, which means that two snapshots before and after a single commit are compared. These methods have achieved high accuracy in detecting refactoring in commits.

\begin{figure}[tb]
    \centering
    \includegraphics[width=0.42\textwidth]{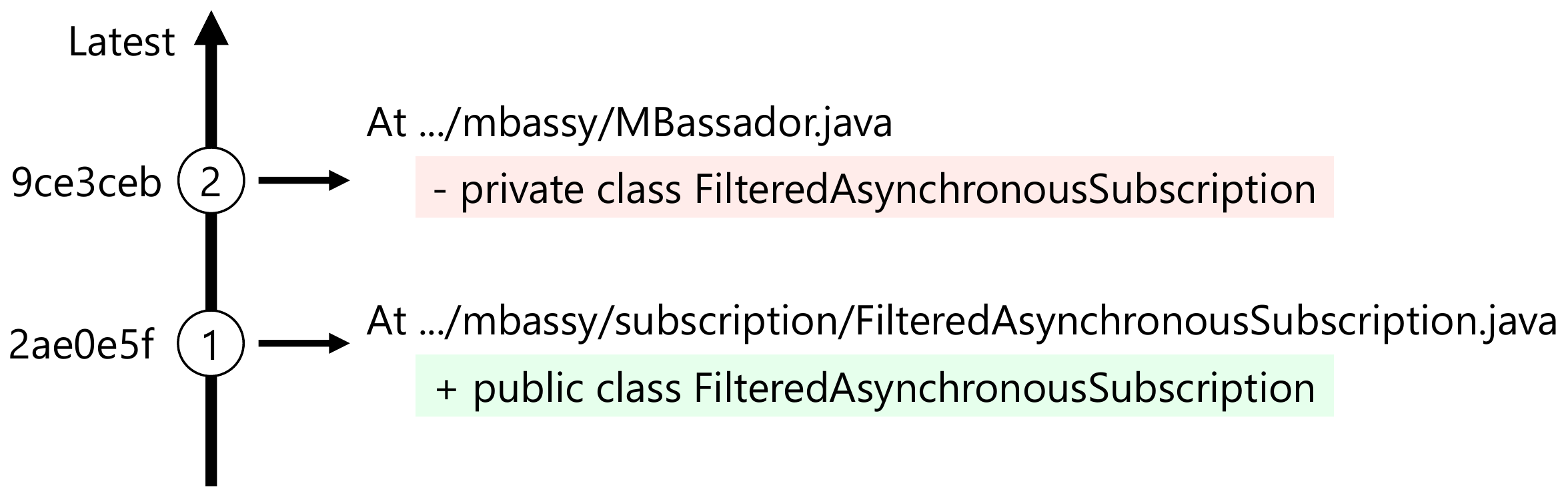}
    \caption{Two commits in \textit{mbassador}.}
    \label{fig :example_coarse_grained_ref}
\end{figure}

However, refactorings that are performed over multiple commits may not be detected. 
The sample history shown in Figure~\ref{fig :example_coarse_grained_ref} consists of two commits extracted from the \textit{mbassador} repository~\cite{mbassador}, where commit \Commit{2ae0e5f} is the parent of commit \Commit{9ce3ceb}.
The intention of the developer, as expressed by these two commits, is to decompose the source file \textit{Mbassador.java}, which contains multiple top-level classes, into multiple source files to ensure that each file contains only one top-level class.
In the first commit, the developer copied the implementation of class \CodeRemoved{FilteredAsynchronousSubscription} in \CodeRemoved{Mbassador.java} to a new file \CodeAdded{FilteredAsynchronousSubscription.java}, and
then she/he  removed that class from the source file \CodeRemoved{Mbassador.java} in the second commit.
Overall, she/he moved a class from \CodeRemoved{Mbassador.java} to a new source file.
A detection based on either of the single commits shown in Figure~\ref{fig :example_coarse_grained_ref} cannot reveal this kind of refactoring because each commit contains only part of the code changes for detecting \Refactoring{Move Class} refactoring.
However, this refactoring can be detected if we consider a coarse-grained commit generated by merging the changes from the two commits.

The existence of refactorings detected only in the granularity of coarse-grained commits suggests that detectors based on single commits may have missed some refactorings.
We conducted an empirical study on 19 open-source Git-based Java repositories to investigate the impact of change granularity in refactoring detection.
To change the granularity of commits, we squashed multiple fine-grained commits into one to form a coarse-grained commit.
The number of fine-grained commits squashed into one coarse-grained commit is referred to as \textit{coarse granularity}.
Refactoring detection is conducted on both fine-grained and coarse-grained commits using the state-of-the-art tool RefactoringMiner~\cite{Tsantalis:TSE:2020:RefactoringMiner2.0,Tsantalis:ICSE:2018:RefactoringMiner}.
If a refactoring type is detected in the coarse-grained commit but not in the fine-grained commits, which formed the coarse one, this refactoring is defined as a \textit{coarse-grained refactoring}~(CGR).

Our results indicate that CGRs are common, and their frequency increases as the granularity becomes coarser. 
The type of refactoring that is most likely to be coarse-grained varies in each repository; however, in general, the \textit{Move}-related refactoring type tends to be CGR.

In summary, our study makes the following contributions:
\begin{itemize}
    \item We propose the definition of CGR.
    \item We evaluate features of CGRs to understand its effect on refactoring detection.
    \item We analyze the reason for the occurrence of CGR.
\end{itemize}

The remainder of this paper is organized as follows.
The next section explains our study design.
Then, we present a preliminary evaluation of 19 open-source projects and the answers to the three research questions in Section~\ref{s:evaluation}.
Finally, in Section~\ref{s:conclusion}, we conclude and state our plans for future work.

\section{Study Design}\label{s:technique}
\begin{figure}[tb]
    \centering
    \includegraphics[width=\linewidth]{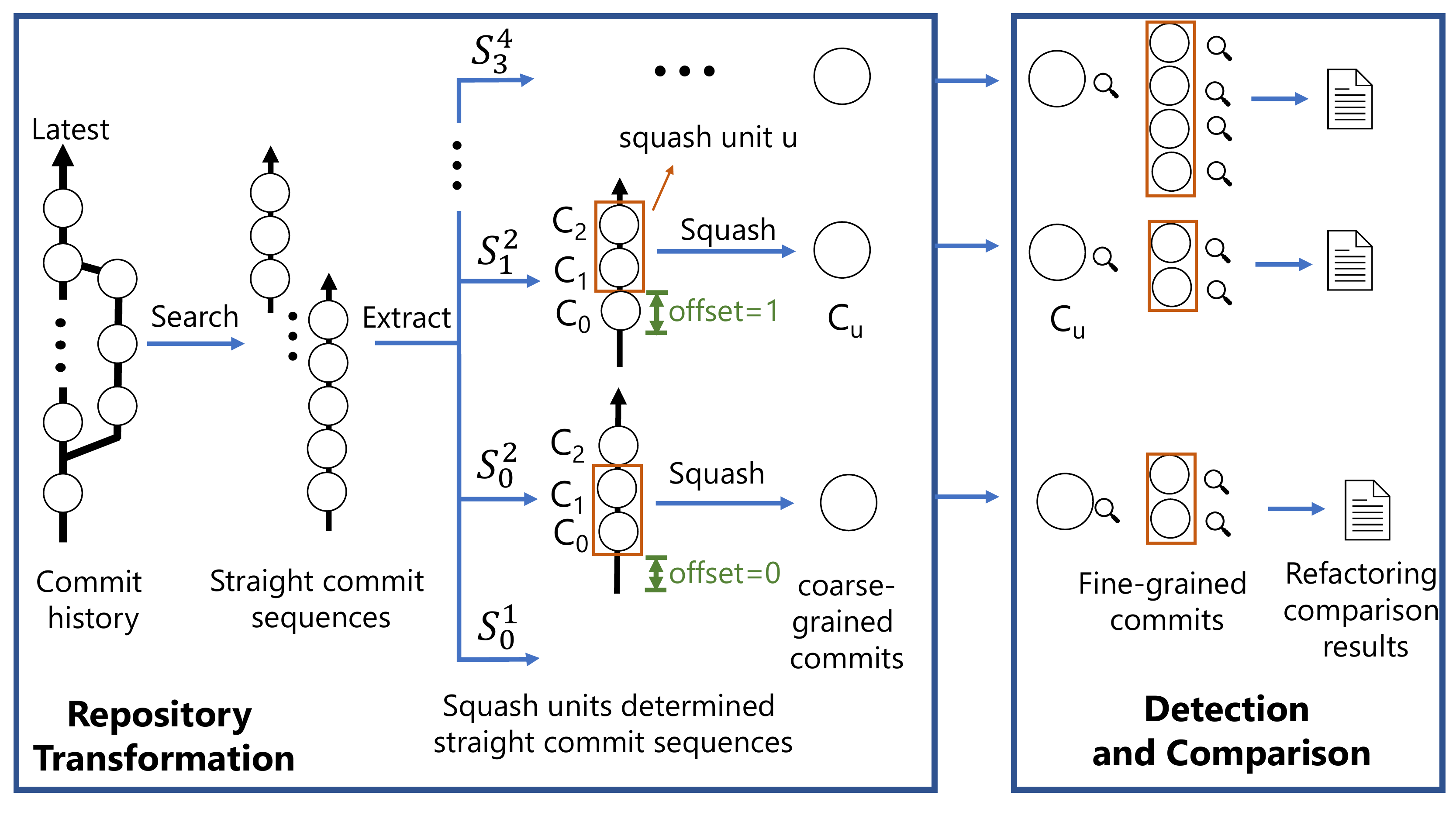}
    \caption{Overview of the study procedure}
    \label{fig: process_overview}
\end{figure}

The overview of our study procedure is shown in Figure~\ref{fig: process_overview}.
Our procedure can be divided into two phases: repository transformation and detection and comparison.
In the repository transformation phase, \textit{squash units} that contain multiple fine-grained commits and can be squashed into coarse-grained ones are extracted from the commit history.
In the detection and comparison phase, refactoring detection is conducted on both fine-grained and coarse-grained commits, and their results are compared.

\subsection{Repository Transformation}\label{ss: Repository Transformation}
In this phase, firstly, the Git-based commit history, as a set of fine-grained commits $H~(\subseteq C)$, is extracted from the given repository, where $C$ is the universal set of commits.
By searching the commit history, we can extract straight commit sequences. Each sequence consists of fine-grained commits that excludes \emph{merge commits}, which have more than one parent, and \emph{branch sources}, which have multiple children.
Merge commits are excluded to avoid duplicate detection of refactoring in the later phase, and
branch sources are excluded for simplicity when extracting squash units.

A \emph{squash unit} $u~(\subseteq C)$ is a set of multiple adjacent fine-grained commits that are squashed into a single coarse-grained commit.
Here, if a commit is the parent or child of another commit, these two commits are considered adjacent. The adjacent commits are shown as circles next to each other in Figure~\ref{fig: process_overview}.
Different strategies labelled $S_o^l$ (for appropriate values of $o$ and $l$) are used to extract squash units from straight commit sequences.
Here, the \emph{granularity level} $l~(\geq 1)$ specifies the size of the squash units, and straight commit sequences are divided into multiple squash units of the specified size.
Because each unit is squashed into one coarse-grained commit, this level expresses the coarse granularity of the coarse-grained commits to be generated.
The granularity level $l = 1$ exactly produces original fine-grained commits.
The \emph{offset} $0 \leq o \leq l - 1$ is the number of commits to be skipped from the beginning of the given straight commit sequence when extracting the squash units to adjust which commits will be merged.
For example, the commit $c_1$ in Figure~\ref{fig: process_overview} is squashed together with $c_{0}$ when strategy $S_0^2$ is used, whereas it is squashed together with $c_{2}$ when strategy $S_1^2$ is used.
For each squash unit $u$, $\squash(u)$ is used to squash all the commits in $u$ into a single coarse-grained commit, which we name $c_u$.

\subsection{Detection and Comparison}
Refactoring detection is conducted on each commit in all extracted squash units and on coarse-grained commits, and the results are compared for each pair of commits.
From commit $c$, a set of refactorings $\detect(c)~(\subseteq R)$ are detected, where $R$ is the universal set of refactorings.
The detection result for one commit contains: 1) the refactoring type, 2) a description of how this refactoring is conducted, and 3) the location where this refactoring is applied in the source code.
Because the location and description of a refactoring may change owing to squashing, we conservatively compared only the type of detected refactorings.
Refactorings detected with invalid locations were excluded.
For a squash unit $u$ and its coarse-grained commit $c_u = \squash(u)$, we judged refactoring $r \in \detect(c_u)$ as coarse-grained if and only if no refactoring of its type $r.\type$ was found in the detected refactorings from each fine-grained commit in $u$.
More specifically, the set of CGRs of $u$ can be explained as
\begin{align*}
  \CGR(u)   &= \{\, r \in \detect(\squash(u)) \mid r.\type \not\in \Types(u) \,\}, \\
  \Types(u) &= \{\, r.\type \mid \exists r \in \detect(c) \wedge c \in u \,\}.
  \addtocounter{equation}{1}\tag{\theequation}
\end{align*}
A squash unit $u$ is regarded as an \emph{effective squash} when at least one CGR is detected from it:
\begin{equation}
    \isEffective(u) = \CGR(u) \neq \emptyset.
\end{equation}
When the coarse granularity is set to $l$, the set of squash units for the repository $H$ is
\begin{equation}
    U_l(H) = \bigcup_{0 \leq o \leq l-1} \Unit_{S_o^l}(H).
\end{equation}
where $\Unit_{S_o^l}(H)$ denotes the squash units extracted from $H$ according to strategy $S_o^l$.

\section{Preliminary Evaluation}\label{s:evaluation}

\subsection{Research Questions}
Our objective in this study is to investigate features of CGRs.
We answer the following research questions (RQs) to better achieve this goal.
\begin{itemize} 
  \item \textbf{\RQ{1}}: \emph{How frequently do CGRs appear because of granularity change?}
  \item \textbf{\RQ{2}}: \emph{Which types of refactorings tend to be coarse-grained?}
  \item \textbf{\RQ{3}}: \emph{What are the reasons for the occurrence of CGRs?}
\end{itemize}

A quantitative analysis is provided for \RQ{1} and \RQ{2}. 
We manually examine the experiment results to present a qualitative explanation for \RQ{3}.

\subsection{Experimental Setup}
We used the Git repository rewriting tool git-stein~\cite{git-stein} to change the granularity and the latest version of RefactoringMiner~(ver.\ 2.2) to detect refactoring in 19 open source Git-based Java repositories.

\subsubsection{Data Collection}
The repositories that we selected are from a dataset collected by Silva \textit{et al.}~\cite{silva2016we},
containing 185 GitHub-hosted Java projects.
Refactorings exist in these projects, some of which have been identified by RefactoringMiner, studied, and confirmed by researchers.
On account of computation time, we chose 19 repositories whose number of commits is no more than 7,000 from the dataset.
To be specific, the number of commits ranges from 342~(\textit{mbassador}) to 6,955~(\textit{redisson}~\cite{redisson}).

\subsection{\RQ{1}: How frequently do CGRs appear because of granularity change?}\label{s:rq1}

\subsubsection{Study Design}
The techniques introduced in Section \ref{s:technique} are applied to the selected repositories to extract squash units, change the granularity of commits, and compare the refactoring detection results to find CGRs.

The frequency of CGRs in the commit history $H$ can be expressed as the ratio of the number of squash units that can generate at least one CGR:
\begin{equation}
    \Frequency(H,l) = \frac{|\{\,u \in U_l(H) \mid \isEffective(u) \,\}|}{|U_l(H)|}.
\end{equation}
We calculate \textit{Frequency} for our dataset when the coarse granularity is set to 2, 3, and 4, respectively. 
\begin{figure}
    \centering
    \includegraphics[width=0.4\textwidth]{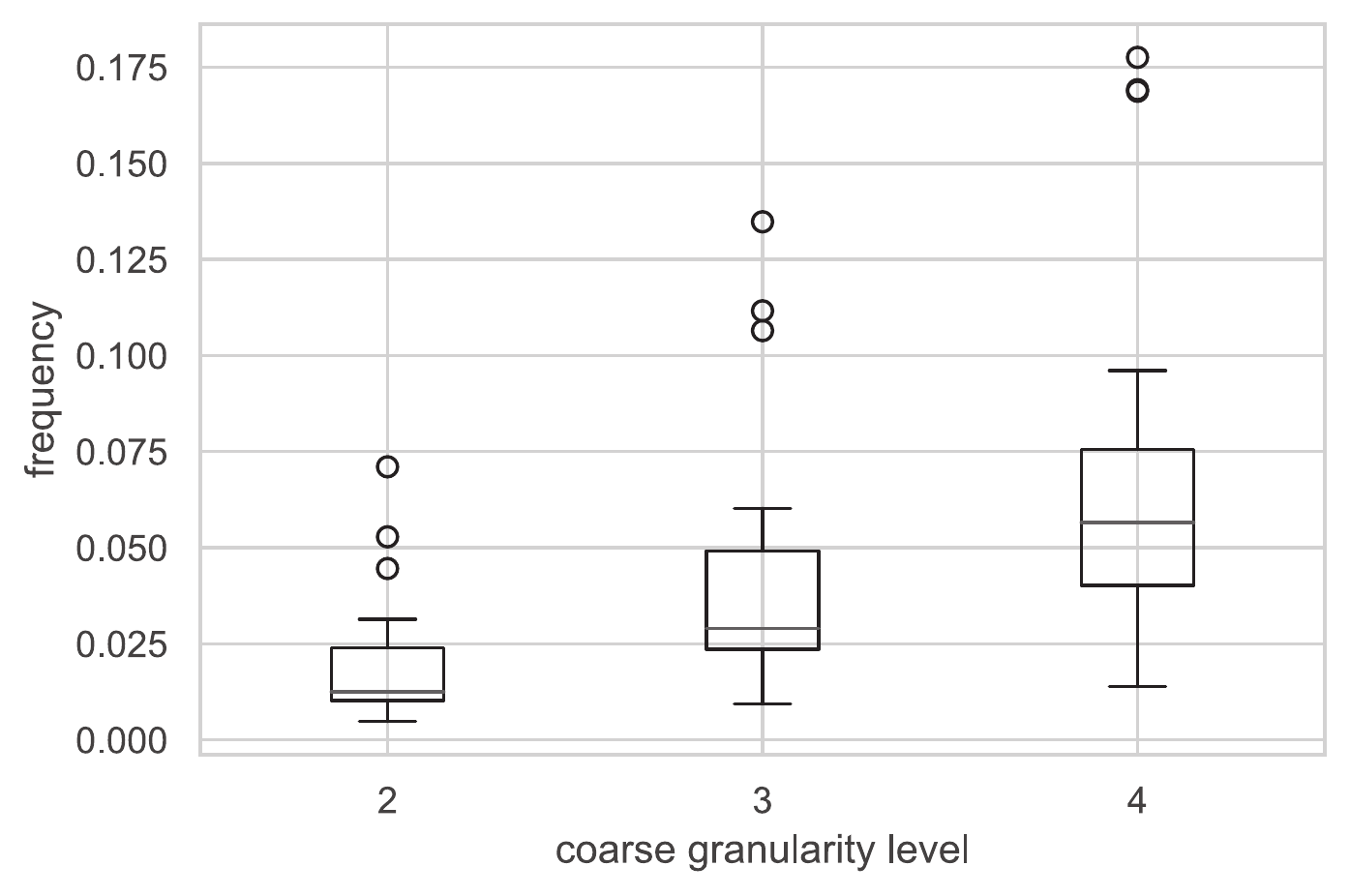}
    \caption{Frequency of CGRs.}
    \label{fig: frequency_boxplot}
\end{figure}
\subsubsection{Results and Discussion}
Figure~\ref{fig: frequency_boxplot} shows box plots of the CGR frequency at different levels of coarse granularity in the 19 repositories.
The minimum values of all three box plots are greater than zero, indicating that CGRs were detected in all the repositories at all levels of coarse granularity.

We can conclude that the CGR is a common phenomenon in refactoring detection.
The highest frequency was observed in the repository \textit{goclipse}~\cite{goclipse}, which was 0.071, 0.135, and 0.178 when the coarse granularity was set to 2, 3, and 4, respectively.
The box plots show that the more the coarse granularity increases, the more the frequency increases in all repositories.
The minimum increase in the frequency when the coarse granularity was changed from 2 to 3 was in the repository \textit{baasbox}~\cite{baasbox}, which increased by 14.1\%, whereas the maximum increase was 331.9\% in \textit{javapoet}~\cite{javapoet}. The average increase for all repositories was 129.4\%. 
When the coarse granularity increases from 3 to 4, a minimum increase of 24.4\% appears in \textit{seyren}~\cite{seyren}, a maximum increase of 147.6\% appears in \textit{mbassador}, and the average increase is 65.6\%.
The average frequencies for all the repositories were 2.0\%, 4.3\%, and 6.9\% when the coarse granularities were 2, 3, and 4, respectively.
The observed tendency of frequency to increase as the coarse granularity increases can be explained as follows.
The CGR detected in the commits with finer granularity may also exist in those with coarser granularity.
In addition, a new CGR may be detected in coarser-grained commits because more code changes are transferred into these commits through the granularity change.

However, we also observed that not all CGRs detected in commits of finer granularity could be detected in a coarser-grained one.
Code changes in other commits may hinder the currently detected CGR when those commits are squashed with the current coarse-grained commit.

\Conclusion{CGR is a common phenomenon in all repositories. 
The average frequencies of CGR for all repositories were 2.0\%, 4.3\%, and 6.9\% when the coarse granularities were 2, 3, and 4, respectively.
CGRs are more frequent when coarse granularity increases.}

\subsection{\RQ{2}: Which types of refactorings tend to be coarse-grained?}\label{s:rq2}
\subsubsection{Study Design}
To investigate this RQ, we calculate the appearance ratio of a specific CGR type at all the three granularity levels.
The ratio expresses the average number of CGRs in one effective squash.
For a certain refactoring type $t$ in commit history $H$, the ratio can be expressed as follows:
\begin{equation}
    \Ratio(t) = \frac{\sum_{2 \leq l \leq 4} |\{\, r \mid \exists u \in U_l(H) \wedge r \in \CGR(u) \wedge r.\type = t \,\}|}%
                     {\sum_{2 \leq l \leq 4} |\{\, u \in U_l(H) \mid \isEffective(u) \,\}|}.
\end{equation}
We calculate the ratio of each type of CGR in our dataset. 

\begin{table}[tb]
    \centering
    \caption{Highest ratio CGR type}\label{table:missing_ratio}
    {\small\begin{tabular}{llc} \hline
    repository         & refactoring type                              & ratio \\ \hline
    jfinal             & \Refactoring{Change Method Access Modifier}   & 0.49 \\
    mbassador          & \Refactoring{Change Class Access Modifier}    & 2.00 \\
	javapoet           & \Refactoring{Replace Variable With Attribute} & 0.80 \\
	jeromq             & \Refactoring{Move Class}                      & 0.19 \\
	seyren             & \Refactoring{Merge Package}                   & 1.00 \\
	retrolambda        & \Refactoring{Push Down Method}                & 1.21 \\
	baasbox            & \Refactoring{Replace Variable With Attribute} & 0.29 \\
	sshj               & \Refactoring{Remove Parameter}                & 0.34 \\
	xabber-android     & \Refactoring{Move Method}                     & 0.30 \\
	android-async-http & \Refactoring{Remove Parameter Modifier}       & 1.40 \\
	giraph             & \Refactoring{Remove Variable Modifier}        & 0.91 \\
	spring-data-rest   & \Refactoring{Move Attribute}                  & 0.19 \\
	blueflood          & \Refactoring{Parameterize Variable}           & 0.08 \\
	HikariCP           & \Refactoring{Move Attribute}                  & 1.82 \\
	redisson           & \Refactoring{Push Down Method}                & 0.12 \\
	goclipse           & \Refactoring{Move Package}                    & 0.05 \\
	atomix             & \Refactoring{Move And Rename Class}           & 0.33 \\
    morphia            & \Refactoring{Move Attribute}                  & 0.71 \\
	PocketHub          & \Refactoring{Move And Rename Class}           & 0.12 \\ \hline
    \end{tabular}}
\end{table}

\subsubsection{Results and Discussion}
The CGR type with the highest ratio for each repository is listed in Table~\ref{table:missing_ratio}. 
Among the 19 repositories, we found that \Refactoring{Change Class Access Modifier} occurs at the highest ratio (2.00) in \textit{mbassador}, and \Refactoring{Move Attribute} in \textit{HikariCP} reaches 1.82.

We find that the CGR type with the highest ratio varies with repositories.
In our dataset, we also find that \textit{Move}-related refactoring types, e.g., \Refactoring{Move Class} and \Refactoring{Move Attribute}, appear most frequently for eight repositories. 
By calculating the average ratio over our dataset for all types of refactorings, we observed that the top three highest-ratio refactoring types were \Refactoring{Move And Rename Class} (0.46\%), \Refactoring{Move Method} (0.34\%), and \Refactoring{Move And Inline Method} (2.9\%).

As a result, we can conclude that \textit{Move}-related refactoring types are most likely to be coarse-grained.
A possible explanation for this is that in \textit{Move}-related refactoring, \textit{Move} on the refactored object is not performed directly but is performed in two steps.
First, an object is copied to the destination and is potentially followed by other changes, e.g., renaming, inline, or no change.
Second, the original object is removed.
These two steps may be included in separate commits.
Another possible reason is that \textit{Move}-related refactoring can be combined with other refactoring, such as \textit{Rename} or \textit{Inline}.
\Conclusion{Considering the average ratio over the whole dataset, the top three types are \Refactoring{Move And Rename Class}, \Refactoring{Move Method}, \Refactoring{Move And Inline Method}.
We conclude that \textit{Move}-related refactoring types are most likely to be coarse-grained.}

\subsection{\RQ{3}: What are the reasons for the occurrence of CGRs?}\label{s:rq3}
\subsubsection{Study Design}
The \emph{git diff} command is used to extract code changes from the fine-grained and coarse-grained commits.
After extraction, we manually compare and analyze the changes and refactorings detected.

\subsubsection{Results and Discussion}
The reasons for the occurrence of CGRs are categorized into two types according to their composition: \textit{Generation} and \textit{Combination}.

\Heading{Generation}
This type of CGR is generated from non-refactoring changes.
The example shown in Figure~\ref{fig :example_coarse_grained_ref} belongs to this type;
the \Refactoring{Move Method} refactoring is generated by two non-refactoring changes:
1) copy the class implementation to a new file
2) remove the origin class.
Another example is in repository \textit{javapoet}.
In the parent commit \Commit{6a3595c}, the attribute \CodeAdded{body} is defined, and the method call \CodeRemoved{methodWriter.write()} is removed.
In child commit \Commit{4ff9adf}, the developer adds method call \CodeAdded{body.write()}. 
In the coarse-grained commit, the above code changes are detected as \Refactoring{Rename Variable with Attribute}; the variable \CodeRemoved{methodWriter} is renamed to attribute \CodeAdded{body}.

\begin{figure}[tb]
    \centering
    \includegraphics[width=0.42\textwidth]{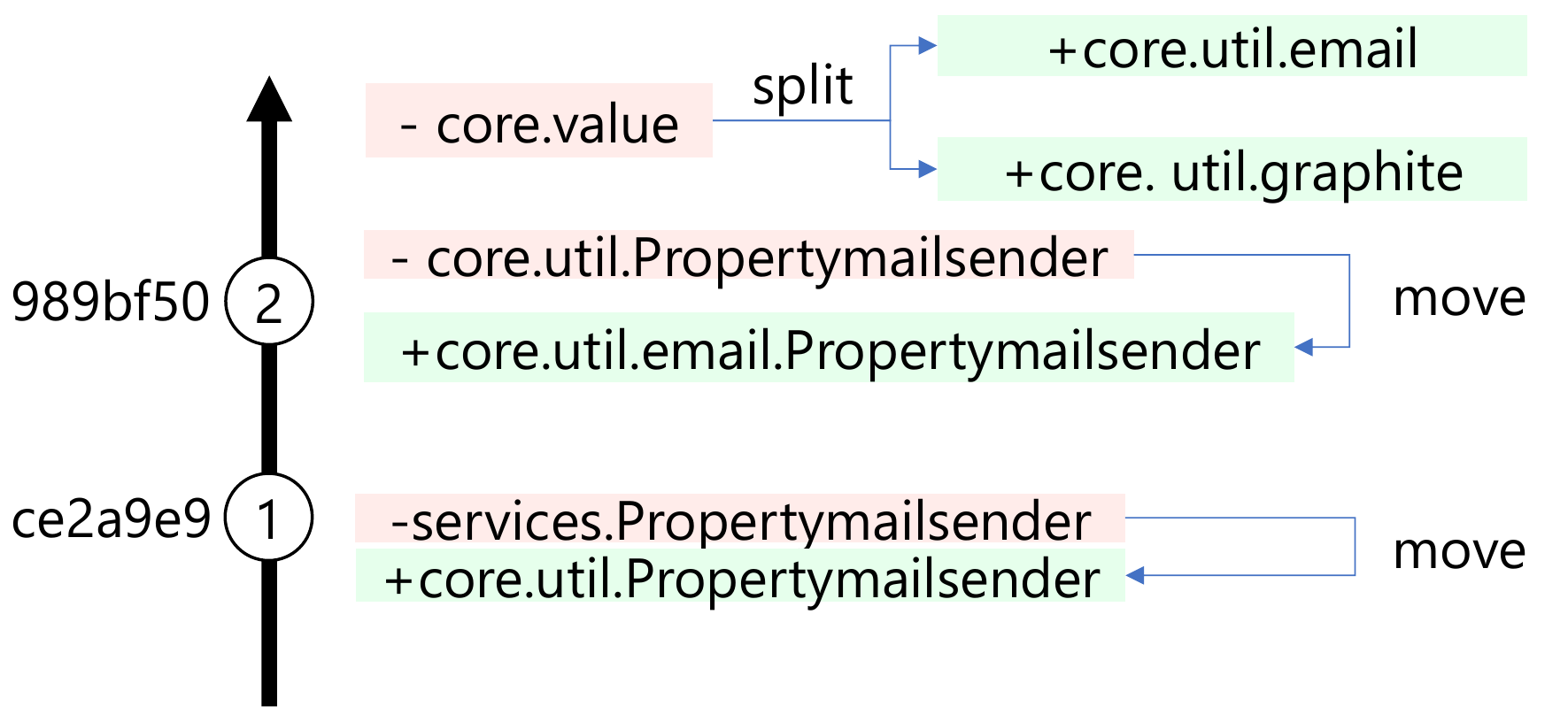}
    \caption{Example of coarse-grained \Refactoring{Merge Package}.}
    \label{fig: seyren_merge_package}
\end{figure}

\Heading{Combination}
In contrast with \textit{Generation}, this type is the combined result of multiple refactorings detected in finer-grained commits.
Figure~\ref{fig: seyren_merge_package} shows an example of this type.
For clarity, only part of the package hierarchy of the repository is shown in the figure. 
In the parent commit \Commit{ce2a9e9}, the developer moves class \CodeRemoved{PropertyMailSender} under package \CodeRemoved{services} to package \CodeAdded{core.util}, which is detected as refactoring \Refactoring{Move Class}.
In child commit \Commit{989bf50}, she/he split package \CodeRemoved{core.value} into \CodeAdded{core.util.email} and another one, and then she/he move class \CodeRemoved{PropertyMailSender} to the package \CodeAdded{core.util.email}, which are detected as \Refactoring{Split Package} and \Refactoring{Move Class}. 
In terms of result, she/he applied \Refactoring{Merge Package} to merge part of the package \CodeRemoved{core.value} and the entire package \CodeRemoved{services} into a new package \CodeAdded{core.util.email}. 

\textit{Generation} type CGRs will influence judgments of whether a module is refactored or not.
We note that this type may also occur because of developers' awareness of refactoring; developers do not realize that the conducted code changes belong to refactoring operations.
Supporting tools to guess developers' manual edits and recognize refactoring activities~\cite{foster2012witchdoctor,10.1145/2568225.2568280} may assist them in development.
Because the \textit{Combination} type may influence type-based refactoring studies, such as investigations on frequently-performed refactoring types, researchers may reconsider their results by covering coarse-grained types.

\Conclusion{We found reasons for two categories.
\textit{Generation} refers to new refactorings generated by non-detected fine-grained ones.
\textit{Combination} is a high-level refactoring combined with detected fine-grained ones.}

\section{Conclusion and Future Work}\label{s:conclusion}
In this study, we investigated the impact of refactoring detection on different granularities of commits in 19 open source Git-based Java repositories.
We observed that it is common for a CGR to occur, and its frequency increases as the granularity becomes coarser.
\textit{Move}-related refactoring types tend to be coarse-grained.
We analyzed the causes of CGR and categorized them into two types according to their composition: \textit{Generation} and \textit{Combination}. 
The studied list of CGR is attached as a supplemental material~\cite{supplemental}.
We suggest that refactoring detectors should cover CGRs.
For future work, we plan to extend the current experiment by comparing different refactoring detection tools on a larger dataset.

\section*{Acknowledgments}
This work was partly supported by the JSPS Grants-in-Aid for Scientific Research JP18K11238, JP21K18302, JP21KK0179, and JP21H04877.


\begin{thebibliography}{24}

    
    \ifx \showCODEN    \undefined \def \showCODEN     #1{\unskip}     \fi
    \ifx \showDOI      \undefined \def \showDOI       #1{#1}\fi
    \ifx \showISBNx    \undefined \def \showISBNx     #1{\unskip}     \fi
    \ifx \showISBNxiii \undefined \def \showISBNxiii  #1{\unskip}     \fi
    \ifx \showISSN     \undefined \def \showISSN      #1{\unskip}     \fi
    \ifx \showLCCN     \undefined \def \showLCCN      #1{\unskip}     \fi
    \ifx \shownote     \undefined \def \shownote      #1{#1}          \fi
    \ifx \showarticletitle \undefined \def \showarticletitle #1{#1}   \fi
    \ifx \showURL      \undefined \def \showURL       {\relax}        \fi
    \providecommand\bibfield[2]{#2}
    \providecommand\bibinfo[2]{#2}
    \providecommand\natexlab[1]{#1}
    \providecommand\showeprint[2][]{arXiv:#2}
    
    \bibitem[\protect\citeauthoryear{??}{mba}{2012}]%
            {mbassador}
     \bibinfo{year}{2012}\natexlab{}.
    \newblock \bibinfo{title}{mbassador}.
    \newblock \bibinfo{howpublished}{https://github.com/bennidi/mbassador}.
    \newblock
    
    
    \bibitem[\protect\citeauthoryear{??}{sey}{2012}]%
            {seyren}
     \bibinfo{year}{2012}\natexlab{}.
    \newblock \bibinfo{title}{seyren}.
    \newblock \bibinfo{howpublished}{https://github.com/scobal/seyren}.
    \newblock
    
    
    \bibitem[\protect\citeauthoryear{??}{baa}{2013}]%
            {baasbox}
     \bibinfo{year}{2013}\natexlab{}.
    \newblock \bibinfo{title}{baasbox}.
    \newblock \bibinfo{howpublished}{https://github.com/baasbox/baasbox}.
    \newblock
    
    
    \bibitem[\protect\citeauthoryear{??}{goc}{2013}]%
            {goclipse}
     \bibinfo{year}{2013}\natexlab{}.
    \newblock \bibinfo{title}{GoClipse}.
    \newblock \bibinfo{howpublished}{https://github.com/GoClipse/goclipse}.
    \newblock
    
    
    \bibitem[\protect\citeauthoryear{??}{jav}{2013}]%
            {javapoet}
     \bibinfo{year}{2013}\natexlab{}.
    \newblock \bibinfo{title}{javapoet}.
    \newblock \bibinfo{howpublished}{https://github.com/square/javapoet}.
    \newblock
    
    
    \bibitem[\protect\citeauthoryear{??}{red}{2014}]%
            {redisson}
     \bibinfo{year}{2014}\natexlab{}.
    \newblock \bibinfo{title}{redisson}.
    \newblock \bibinfo{howpublished}{https://github.com/redisson/redisson}.
    \newblock
    
    
    \bibitem[\protect\citeauthoryear{Bavota, De~Carluccio, De~Lucia, Di~Penta,
      Oliveto, and Strollo}{Bavota et~al\mbox{.}}{2012}]%
            {6392107}
    \bibfield{author}{\bibinfo{person}{Gabriele Bavota},
      \bibinfo{person}{Bernardino De~Carluccio}, \bibinfo{person}{Andrea De~Lucia},
      \bibinfo{person}{Massimiliano Di~Penta}, \bibinfo{person}{Rocco Oliveto},
      {and} \bibinfo{person}{Orazio Strollo}.} \bibinfo{year}{2012}\natexlab{}.
    \newblock \showarticletitle{When Does a Refactoring Induce Bugs? An Empirical
      Study}. In \bibinfo{booktitle}{\emph{Proceedings of the 12th IEEE
      International Working Conference on Source Code Analysis and Manipulation
      (SCAM 2012)}}. \bibinfo{pages}{104--113}.
    \newblock
    
    
    \bibitem[\protect\citeauthoryear{Ch\'{a}vez, Ferreira, Fernandes, Cedrim, and
      Garcia}{Ch\'{a}vez et~al\mbox{.}}{2017}]%
            {10.1145/3131151.3131171}
    \bibfield{author}{\bibinfo{person}{Alexander Ch\'{a}vez},
      \bibinfo{person}{Isabella Ferreira}, \bibinfo{person}{Eduardo Fernandes},
      \bibinfo{person}{Diego Cedrim}, {and} \bibinfo{person}{Alessandro Garcia}.}
      \bibinfo{year}{2017}\natexlab{}.
    \newblock \showarticletitle{How Does Refactoring Affect Internal Quality
      Attributes? A Multi-Project Study}. In \bibinfo{booktitle}{\emph{Proceedings
      of the 31st Brazilian Symposium on Software Engineering (SBES 2017)}}.
      \bibinfo{pages}{74–83}.
    \newblock
    
    
    \bibitem[\protect\citeauthoryear{Chen and Hayashi}{Chen and Hayashi}{2022}]%
            {supplemental}
    \bibfield{author}{\bibinfo{person}{Lei Chen} {and} \bibinfo{person}{Shinpei
      Hayashi}.} \bibinfo{year}{2022}\natexlab{}.
    \newblock \bibinfo{title}{Appendix of ``Impact of Change Granularity in
      Refactoring Detection''}.
    \newblock \bibinfo{howpublished}{https://doi.org/10.5281/zenodo.6399250}.
    \newblock
    
    
    \bibitem[\protect\citeauthoryear{Dig, Comertoglu, Marinov, and Johnson}{Dig
      et~al\mbox{.}}{2006}]%
            {dig2006automated}
    \bibfield{author}{\bibinfo{person}{Danny Dig}, \bibinfo{person}{Can
      Comertoglu}, \bibinfo{person}{Darko Marinov}, {and} \bibinfo{person}{Ralph
      Johnson}.} \bibinfo{year}{2006}\natexlab{}.
    \newblock \showarticletitle{Automated detection of refactorings in evolving
      components}. In \bibinfo{booktitle}{\emph{Proceedings of the 20th European
      Conference on Object-Oriented Programming (ECOOP 2006)}}.
      \bibinfo{pages}{404--428}.
    \newblock
    
    
    \bibitem[\protect\citeauthoryear{Fernandes, Chávez, Garcia, Ferreira, Cedrim,
      Sousa, and Oizumi}{Fernandes et~al\mbox{.}}{2020}]%
            {FERNANDES2020106347}
    \bibfield{author}{\bibinfo{person}{Eduardo Fernandes},
      \bibinfo{person}{Alexander Chávez}, \bibinfo{person}{Alessandro Garcia},
      \bibinfo{person}{Isabella Ferreira}, \bibinfo{person}{Diego Cedrim},
      \bibinfo{person}{Leonardo Sousa}, {and} \bibinfo{person}{Willian Oizumi}.}
      \bibinfo{year}{2020}\natexlab{}.
    \newblock \showarticletitle{Refactoring effect on internal quality attributes:
      What haven't they told you yet?}
    \newblock \bibinfo{journal}{\emph{Information and Software Technology}}
      \bibinfo{volume}{126} (\bibinfo{year}{2020}), \bibinfo{pages}{106347}.
    \newblock
    
    
    \bibitem[\protect\citeauthoryear{Foster, Griswold, and Lerner}{Foster
      et~al\mbox{.}}{2012}]%
            {foster2012witchdoctor}
    \bibfield{author}{\bibinfo{person}{Stephen~R. Foster},
      \bibinfo{person}{William~G. Griswold}, {and} \bibinfo{person}{Sorin Lerner}.}
      \bibinfo{year}{2012}\natexlab{}.
    \newblock \showarticletitle{WitchDoctor: IDE support for real-time
      auto-completion of refactorings}. In \bibinfo{booktitle}{\emph{Proceedings of
      the 34th International Conference on Software Engineering (ICSE 2012)}}.
      \bibinfo{pages}{222--232}.
    \newblock
    
    
    \bibitem[\protect\citeauthoryear{Ge and Murphy-Hill}{Ge and
      Murphy-Hill}{2014}]%
            {10.1145/2568225.2568280}
    \bibfield{author}{\bibinfo{person}{Xi Ge} {and} \bibinfo{person}{Emerson
      Murphy-Hill}.} \bibinfo{year}{2014}\natexlab{}.
    \newblock \showarticletitle{Manual Refactoring Changes with Automated
      Refactoring Validation}. In \bibinfo{booktitle}{\emph{Proceedings of the 36th
      International Conference on Software Engineering (ICSE 2014)}}.
      \bibinfo{pages}{1095–1105}.
    \newblock
    
    
    \bibitem[\protect\citeauthoryear{Hayashi}{Hayashi}{2018}]%
            {git-stein}
    \bibfield{author}{\bibinfo{person}{Shinpei Hayashi}.}
      \bibinfo{year}{2018}\natexlab{}.
    \newblock \bibinfo{title}{git-stein}.
    \newblock \bibinfo{howpublished}{https://github.com/sh5i/git-stein}.
    \newblock
    
    
    \bibitem[\protect\citeauthoryear{Kim, Gee, Loh, and Rachatasumrit}{Kim
      et~al\mbox{.}}{2010}]%
            {10.1145/1882291.1882353}
    \bibfield{author}{\bibinfo{person}{Miryung Kim}, \bibinfo{person}{Matthew Gee},
      \bibinfo{person}{Alex Loh}, {and} \bibinfo{person}{Napol Rachatasumrit}.}
      \bibinfo{year}{2010}\natexlab{}.
    \newblock \showarticletitle{Ref-Finder: A Refactoring Reconstruction Tool Based
      on Logic Query Templates}. In \bibinfo{booktitle}{\emph{Proceedings of the
      18th ACM SIGSOFT International Symposium on Foundations of Software
      Engineering (FSE 2010)}}. \bibinfo{pages}{371–372}.
    \newblock
    \showISBNx{9781605587912}
    
    
    \bibitem[\protect\citeauthoryear{Kim, Zimmermann, and Nagappan}{Kim
      et~al\mbox{.}}{2012}]%
            {kim2012field}
    \bibfield{author}{\bibinfo{person}{Miryung Kim}, \bibinfo{person}{Thomas
      Zimmermann}, {and} \bibinfo{person}{Nachiappan Nagappan}.}
      \bibinfo{year}{2012}\natexlab{}.
    \newblock \showarticletitle{A field study of refactoring challenges and
      benefits}. In \bibinfo{booktitle}{\emph{Proceedings of the ACM SIGSOFT 20th
      International Symposium on the Foundations of Software Engineering (FSE
      2012)}}. \bibinfo{pages}{1--11}.
    \newblock
    
    
    \bibitem[\protect\citeauthoryear{Kim, Zimmermann, and Nagappan}{Kim
      et~al\mbox{.}}{2014}]%
            {kim2014empirical}
    \bibfield{author}{\bibinfo{person}{Miryung Kim}, \bibinfo{person}{Thomas
      Zimmermann}, {and} \bibinfo{person}{Nachiappan Nagappan}.}
      \bibinfo{year}{2014}\natexlab{}.
    \newblock \showarticletitle{An empirical study of refactoring challenges and
      benefits at microsoft}.
    \newblock \bibinfo{journal}{\emph{IEEE Transactions on Software Engineering}}
      \bibinfo{volume}{40}, \bibinfo{number}{7} (\bibinfo{year}{2014}),
      \bibinfo{pages}{633--649}.
    \newblock
    
    
    \bibitem[\protect\citeauthoryear{Prete, Rachatasumrit, Sudan, and Kim}{Prete
      et~al\mbox{.}}{2010}]%
            {prete-icsm2010}
    \bibfield{author}{\bibinfo{person}{Kyle Prete}, \bibinfo{person}{Napol
      Rachatasumrit}, \bibinfo{person}{Nikita Sudan}, {and}
      \bibinfo{person}{Miryung Kim}.} \bibinfo{year}{2010}\natexlab{}.
    \newblock \showarticletitle{Template-based Reconstruction of Complex
      Refactorings}. In \bibinfo{booktitle}{\emph{Proceedings of the 26th IEEE
      International Conference on Software Maintenance (ICSM 2010)}}.
      \bibinfo{pages}{1--10}.
    \newblock
    
    
    \bibitem[\protect\citeauthoryear{Silva, da~Silva, Santos, Terra, and
      Valente}{Silva et~al\mbox{.}}{2021}]%
            {refdiff20}
    \bibfield{author}{\bibinfo{person}{Danilo Silva}, \bibinfo{person}{João~Paulo
      da Silva}, \bibinfo{person}{Gustavo Santos}, \bibinfo{person}{Ricardo Terra},
      {and} \bibinfo{person}{Marco~Tulio Valente}.}
      \bibinfo{year}{2021}\natexlab{}.
    \newblock \showarticletitle{{RefDiff 2.0}: A Multi-Language Refactoring
      Detection Tool}.
    \newblock \bibinfo{journal}{\emph{IEEE Transactions on Software Engineering}}
      \bibinfo{volume}{47}, \bibinfo{number}{12} (\bibinfo{year}{2021}),
      \bibinfo{pages}{2786--2802}.
    \newblock
    
    
    \bibitem[\protect\citeauthoryear{Silva, Tsantalis, and Valente}{Silva
      et~al\mbox{.}}{2016}]%
            {silva2016we}
    \bibfield{author}{\bibinfo{person}{Danilo Silva}, \bibinfo{person}{Nikolaos
      Tsantalis}, {and} \bibinfo{person}{Marco~Tulio Valente}.}
      \bibinfo{year}{2016}\natexlab{}.
    \newblock \showarticletitle{Why we refactor? Confessions of GitHub
      contributors}. In \bibinfo{booktitle}{\emph{Proceedings of the 24th ACM
      SIGSOFT International Symposium on Foundations of Software Engineering (FSE
      2016)}}. \bibinfo{pages}{858--870}.
    \newblock
    
    
    \bibitem[\protect\citeauthoryear{Silva and Valente}{Silva and Valente}{2017}]%
            {silva2017refdiff}
    \bibfield{author}{\bibinfo{person}{Danilo Silva} {and}
      \bibinfo{person}{Marco~Tulio Valente}.} \bibinfo{year}{2017}\natexlab{}.
    \newblock \showarticletitle{RefDiff: Detecting Refactorings in Version
      Histories}. In \bibinfo{booktitle}{\emph{Proceedings of the 14th IEEE/ACM
      International Conference on Mining Software Repositories (MSR 2017)}}.
      \bibinfo{pages}{269--279}.
    \newblock
    
    
    \bibitem[\protect\citeauthoryear{Tsantalis, Ketkar, and Dig}{Tsantalis
      et~al\mbox{.}}{2020}]%
            {Tsantalis:TSE:2020:RefactoringMiner2.0}
    \bibfield{author}{\bibinfo{person}{Nikolaos Tsantalis}, \bibinfo{person}{Ameya
      Ketkar}, {and} \bibinfo{person}{Danny Dig}.} \bibinfo{year}{2020}\natexlab{}.
    \newblock \showarticletitle{RefactoringMiner 2.0}.
    \newblock \bibinfo{journal}{\emph{IEEE Transactions on Software Engineering}}
      \bibinfo{volume}{48}, \bibinfo{number}{3} (\bibinfo{year}{2020}),
      \bibinfo{pages}{930--950}.
    \newblock
    
    
    \bibitem[\protect\citeauthoryear{Tsantalis, Mansouri, Eshkevari, Mazinanian,
      and Dig}{Tsantalis et~al\mbox{.}}{2018}]%
            {Tsantalis:ICSE:2018:RefactoringMiner}
    \bibfield{author}{\bibinfo{person}{Nikolaos Tsantalis}, \bibinfo{person}{Matin
      Mansouri}, \bibinfo{person}{Laleh~M. Eshkevari}, \bibinfo{person}{Davood
      Mazinanian}, {and} \bibinfo{person}{Danny Dig}.}
      \bibinfo{year}{2018}\natexlab{}.
    \newblock \showarticletitle{Accurate and Efficient Refactoring Detection in
      Commit History}. In \bibinfo{booktitle}{\emph{Proceedings of the 40th
      International Conference on Software Engineering (ICSE 2018)}}.
      \bibinfo{pages}{483--494}.
    \newblock
    
    
    \bibitem[\protect\citeauthoryear{Wei{\ss}gerber and Diehl}{Wei{\ss}gerber and
      Diehl}{2006}]%
            {weissgerber-ase2006}
    \bibfield{author}{\bibinfo{person}{Peter Wei{\ss}gerber} {and}
      \bibinfo{person}{Stephan Diehl}.} \bibinfo{year}{2006}\natexlab{}.
    \newblock \showarticletitle{Identifying Refactorings from Source-Code Changes}.
      In \bibinfo{booktitle}{\emph{Proceedings of the 21st International Conference
      on Automated Software Engineering (ASE 2006)}}. \bibinfo{pages}{231--240}.
    \newblock
    
    
    \end{thebibliography}


\end{document}